\documentclass[12pt]{article}
\usepackage{amssymb,amsmath}

\newtheorem{theorem}{Theorem}[subsection]
\newtheorem{lemma}[theorem]{Lemma}

\newenvironment{proof}[1][Proof]{\begin{trivlist}
\item[\hskip \labelsep {\bfseries
#1}]}{\end{trivlist}}

 \newcommand{\qed}{\nobreak
\ifvmode \relax \else
\ifdim\lastskip<1.5em \hskip-\lastskip
\hskip1.5em plus0em minus0.5em \fi \nobreak \vrule height0.75em width0.5em
depth0.25em\fi}
\newcommand{\E}{\textup{\bf E}}

\begin{document}

\begin{centering}

{\bf \Large Canonical Transformations and Hamiltonian}

{\bf \Large \vspace{.5cm} Evolutionary Systems}

\renewcommand{\thefootnote}{\fnsymbol{footnote}}
\vspace{1.0cm}
{\large Samer Ashhab}\\

\vspace{.5cm}

Department of Mathematics, University of New Orleans, New Orleans, \\
LA 70148.\\
\it E-mail: salashha@uno.edu\\
\rm
\vspace{.5cm}


\begin{abstract}

In many Lagrangian field theories one has a Poisson bracket defined on the space of local functionals. We find necessary and sufficient conditions
for a transformation on the space of local functionals to be canonical in three different cases. These three cases depend on the specific dimensions of the vector bundle of the theory and the associated Hamiltonian differential operator. We also show how a canonical transformation transforms a Hamiltonian evolutionary system and its conservation laws. Finally we illustrate these ideas with three examples.

\end{abstract}

\end{centering}

\noindent
Keywords: Hamiltonian evolutionary systems, Poisson bracket, jet bundle. \\
AMS Subject Classification; Primary: 37K05. Secondary: 53Z05, 35Q99.

\section{Introduction}
In Hamiltonian evolutionary systems one has a Poisson bracket. This Poisson bracket is defined on the space of local functionals of a fiber bundle.
The system is characterized by a Hamiltonian which is a local functional and the associated Poisson bracket. The Poisson bracket in turn is characterized by a {\em Hamiltonian} differential operator (see the discussion in the next section).
The solutions to the evolutionary system are sections of the fiber bundle.
Generally an evolutionary system has the form $$u_t=\mathcal D \delta \mathcal H$$
where $\mathcal D$ is the associated {\em Hamiltonian} differential operator,
$\delta$ the variational derivative and $\mathcal H$ the Hamiltonian of the system. The variational derivative
of a local functional $\mathcal P = \int_M P \nu$ has the same components as $\E(P \nu)$ where $\E$ is the Euler-Lagrange
operator and $P$ is a local function representing the local functional $\mathcal P$, while $\nu$ is a volume element on the base manifold $M$. A good reference
on evolutionary systems and their theory can be found in \cite {D91} and \cite {O86} for example.
The reader may consult \cite {KV98} as well. \\

An action on the bundle
induces an action/transformation on the space of local functionals. If this transformation preserves the Poisson
bracket we call it a canonical transformation. This terminology was used in \cite{AF03} and is
in anology to the one used in \cite{MR94} for the case of symplectic manifolds.
In \cite{AF03} canonical transformations
were studied in the case the Poisson bracket is defined by a differential operator of order {\em zero}. In
this paper we study canonical transofrmations in the case of higher order differential operators.
We work with vector bundles with $m$-dimensional fibers and $n$-dimensional base space. We consider special cases of higher order differential
operators which we find applications for and intend to broaden our study in future work. \\

A canonical transformation will transform the evolutionary system, its Hamiltonian, and its conservation laws where
a new system is obtained. We will show that the conditions for a canonical transformation are rather
restrictive in some cases. We illustrate by a few examples at the end of the paper. \\

It would be interesting to study the sh-Lie algebra structure and reduction as was done in \cite{AF03}
and \cite{BFLS98}. However we leave this for a possible future work.

\section{Background material}
Let $\pi :E \to M$ be a vector bundle of dimension $m+n$ and with a base space an $n$-dimensional manifold $M$. Let
$J^\infty E$ be the infinite jet bundle of $E.$ The restriction of the infinite
jet bundle over  an appropriate open set $U\subset M$ is trivial with fiber an
infinite dimensional vector space $V^\infty$.  The bundle \begin{eqnarray*}
\pi^\infty : J^\infty E_U=U\times V^\infty \rightarrow U \end{eqnarray*} then
has induced coordinates given by \begin{eqnarray*}
(x^i,u^a,u^a_i,u^a_{i_1i_2},\dots,). \end{eqnarray*} We use multi-index notation
and the summation convention throughout the paper. If $j^{\infty}\phi$ is the
section of $J^{\infty}E$ induced by a section $\phi$ of the bundle $E$, then
$u^a\circ j^{\infty}\phi=u^a\circ \phi$ and $$u^a_I\circ j^{\infty}\phi=
(\partial_{i_1}\partial_{i_2}...\partial_{i_r})(u^a\circ j^{\infty}\phi)$$
where $r$ is the order of the symmetric multi-index
$I=\{i_1,i_2,...,i_r\}$, with the convention that, for $r=0$, there are no
derivatives. For more details see \cite{A96} and \cite{KV98}.

Let $Loc_E$ denote the algebra of local functions where a local function on
$J^\infty E$ is defined to be the pull-back of a smooth function on some finite
jet bundle $J^p E$ via the projection from $J^\infty E$ to $J^p E$. Let
$Loc_E^0$ denote the subalgebra of $Loc_E$ such that $P \in Loc_E^0$ iff
$(j^\infty \phi)^* P$ has compact support for all $\phi \in \Gamma E$ with
compact support and where $\Gamma E$ denotes the set of sections of the bundle $E
\to M$. The de Rham complex of differential forms $\Omega^*(J^{\infty}E,d)$ on
$J^{\infty}E$ possesses a differential ideal, the ideal ${ C}$ of contact forms
$\theta$ which satisfy $(j^{\infty}\phi)^* \theta=0$ for all sections $\phi$
with compact support. This ideal is generated by the contact one-forms, which
in local coordinates assume the form $\theta^a_J=du^a_J-u^a_{iJ}dx^i$.

Now let $C_0$ denote the set of contact one-forms of {\em order zero}. Contact
one-forms of order zero satisfy $(j^{1}\phi)^*(\theta)=0$ and, in local
coordinates, they assume the form $\theta^a= du^a-u^a_idx^i$. Notice that both
$C_0$ and $\Omega ^{n,1} = \Omega ^{n,1}(J^ \infty E)$ are modules over
$Loc_E$. Let $\Omega ^{n,1}_0$ denote the subspace of $\Omega ^{n,1}$ which is
locally generated by the forms $\{(\theta ^a \wedge d^nx)\}$ over $Loc_E$. Let $\nu$ denote a
volume element on $M$  and notice that in local coordinates $\nu$ takes the form
$\nu = f d^nx = f dx^1 \wedge dx^2 \wedge ... \wedge dx^n$ for some function
$f: U \to {\mathbf R}$ and $U$ is a subset of $M$ on which the $x^i$'s are
defined. \bigskip

Define the operator $D_i$ (total derivative) by $\displaystyle D_i =
\frac{\partial}{\partial x^i} + u^a_{iJ}\frac{\partial}
{\partial u^a_J}$ (recall we assume the summation convention, i.e.,
the sum is over all $a$ and multi-index $J$). For $I=\{i_1i_2\cdots i_k\}$ where $k>0$, $D_I$ is defined by $D_I=D_{i_1}\circ D_{i_2} \circ \cdots \circ D_{i_k}$. If $I$ is empty $I=\{\}$ then $D_I$ is just multiplication by 1. We also define $(-D)_I = (-1)^{|I|} D_I$. Recall that the
Euler-Lagrange operator
maps
$\Omega ^{n,0}(J^\infty E)$ into $\Omega ^{n,1}_0(J^\infty E)$ and is
defined
by $$\E (P \nu)=\E _a(P)(\theta ^a \wedge \nu)$$ where $P \in Loc_E, \nu$
is a
volume
form on the base manifold $M$,
and the
components $\E_a(P)$ are given by $$\E_a(P)=(-D)_I(\frac{\partial P}
{\partial u^a_I}).$$ For simplicity of notation we may use $\E(P)$ for $\E
(P\nu)$.
We will also use $\tilde{D}_i$
for $\displaystyle \frac{\partial}{\partial \tilde{x}^i} +
\tilde{u}^a_{iJ}\frac{\partial}{\partial \tilde{u}^a_J}$
and $\tilde{\E}_a(P)$ for
$\displaystyle (-\tilde{D})_I(\frac {\partial P} {\partial
\tilde{u}^a_I})$
so that $\E(P)
= \tilde{\E}_a(P) (\tilde{\theta} ^a \wedge \nu)$ in the
$(\tilde{x}^{\mu},\tilde{u}^a)$ coordinate system. \\

Now let $\mathcal F$ be the space of functionals where $\mathcal P \in
\mathcal F$ iff $\displaystyle \mathcal P=\int_M P \nu$ for some $P \in Loc_E^0$.
Let $\mathcal D$ be a differential operator which has components $\mathcal D ^{ab}=\omega ^{abI} D_I$ where $\omega ^{abI} \in Loc_E$ and $D_I$ is a combination of the total derivatives as determined by the multi-index $I$, i.e., $D_I$ is a composite of the form $D_{i_1}\circ D_{i_2} \circ \cdots \circ D_{i_k}$.
Define a Poisson bracket on the space of local functionals $\mathcal F$ by $$\{\mathcal P,\mathcal Q\}(\phi)=
\int_M [\E(P)\mathcal D (\E(Q))\circ j\phi] \nu,$$
where $\phi \in \Gamma E$, $\nu$ is a volume form on $\displaystyle M, \mathcal P=
\int_M P \nu, \mathcal Q=\int_M Q \nu,$ and $P, Q \in Loc_E^0$. In this expression one may represent $\mathcal D$ as a matrix differential operator and $\E$ as a
row/column vector (as appropriate) consisting of the components $\E_a$.
We assume that $\mathcal D$ is {\em Hamiltonian} so that we have a genuine Poisson bracket that is antisymmetric and satisfies the Jacobi identity (e.g. see \cite{O86}).
Using local coordinates $(x^\mu,u^a_I)$ on $J^\infty
E$, observe that for $\phi \in \Gamma E$ such that the support of $\phi$
lies in the domain $\Omega$ of some chart
$x$ of $M$, one has
$$\{\mathcal P,\mathcal Q\}(\phi) =
\int_{x(\Omega)} ([\E_a(P)\mathcal D^{ab} (\E_b(Q))] \circ j\phi \circ x^{-1})
(x^{-1})^*(\nu)$$
where $x^{-1}$ is the inverse of $x=(x^\mu)$. \\

The functions $P$ and $Q$ in our definition of the Poisson bracket (of local
functionals) are representatives of $\mathcal P$ and $\mathcal Q$ respectively,
since generally these are not unique. In fact $\mathcal F \simeq H^n_c(J^\infty
E)$, where $H^n_c(J^\infty E) = \Omega ^{n,0}_{c}(J^\infty E)/(\textup{im}
d_H\bigcap\Omega ^{n,0}_{c}(J^\infty E))$ and im$d_H$ is the image of the
differential $d_H$ defined by $d_H = dx^i D_i$. We refer the reader to \cite{AF03} for more details and for the notation.
The interested reader may also consult \cite{A96} and \cite{BT82} for more on the de Rham complex and
its cohomology. \\

Let $\psi:E\rightarrow E$ be an automorphism, sending fibers to fibers, and let
$\psi _M:M\rightarrow M$ be the induced diffeomorphism of $M$. Notice that
$\psi$ induces an automorphism $j\psi: J^{\infty}E \rightarrow J^{\infty}E$
where $$(j\psi)((j^\infty \phi)(p))=j(\psi \circ \phi \circ \psi ^{-1}_M)(\psi
_M(p)),$$ for all $\phi \in \Gamma E$ and all $p$ in the domain of $\phi$.  In
these coordinates the independent variables transform via $\tilde{x}^\mu = \psi
_M^\mu(x^\nu)$. Local coordinate representatives of $\psi _M$ and $j\psi$ may
be described in terms of charts $(\Omega,x)$ and $(\tilde{\Omega},\tilde{x})$
of $M$, and induced charts $((\pi ^\infty)^{-1} (\Omega),(x^{\mu},u^a_I))$ and
$((\pi ^\infty)^{-1} (\tilde{\Omega}),(\tilde{x}^{\mu},\tilde{u}^a_I))$ of $J^
\infty E$. \\

Observe that the total derivatives satisfy 
\begin{equation} \label{der} D_i (F\circ j\psi)= ((\tilde D_j F)\circ j\psi) D_i \psi _M ^j
\end{equation}
where $\tilde{x}^j = \psi_M^j(x).$ \\ \\
\noindent
{\bf Example} Let $M={\bf R}^2, E={\bf R}^2\times{\bf R}^1$ and 
consider
the transformation $\psi$ defined by $\tilde {u} =xu+3yu^2, 
\tilde x=x\cos\theta+y\sin\theta,
\tilde y=-x\sin\theta+y\cos\theta$. Now let $F=\tilde u_{\tilde x}$ then $F\circ 
j\psi=(xu_x+u+6yuu_x)\cos\theta
+(xu_y+3u^2+6yuu_y)\sin\theta$ so that $D_x(F\circ j\psi)=(xu_{xx}+2u_x
+6yu_x^2+6yuu_{xx})\cos\theta + (xu_{yx}+u_y+6uu_x+6yu_xu_y+6yuu_{yx})
\sin\theta$. On the other hand the right-hand side of equation ($\ref {der}$) 
yields $[\tilde u_{\tilde{x}\tilde{x}}\cos\theta+\tilde u_{\tilde{x}\tilde{y}}(-\sin\theta)]\circ j\psi$ 
which when evaluated and simplified yields the same expression as above. \\

\noindent {\bf Remark} For simplicity we may skip writing the tilde's, so in the above example one simply writes $F=u_x$ instead of $F=\tilde u_{\tilde x}$, ...etc.

\section{Canonical transformations of the Poisson structure}
Let $L:J^{\infty}E \rightarrow {\bf R}$ be a Lagrangian in $Loc_E$ (generally
we will assume that any element of $Loc_E$ is a Lagrangian). Let $\hat{L}
= L\circ (x^{\mu},u^a_I)^{-1}$ and let $\tilde{L} = L\circ
(\tilde{x}^{\mu},\tilde{u}^a_I)^{-1}$. Then, in local coordinates, $\tilde{L}$
is related to $\hat{L}$
by the equation $$(\tilde{L} \circ j\bar{\psi}) \textup{det}(J) = \hat{L},$$
where $j\bar{\psi} = (\tilde{x}^{\nu},\tilde{u}^b_K) \circ
(x^{\mu}, u^a_I)^{-1}$ and $J$ is the Jacobian matrix of the transformation
$\psi _M= \tilde{x}^\nu \circ (x^\mu)^{-1}$.
With abuse of notation we may assume coordinates and charts are the same and
write $\tilde{x}^\nu = \psi _M(x^\mu)$.
For simplicity, we have also assumed that
$\psi _M$ is orientation-preserving. In this case the functional
$$\tilde{\mathcal L}=\int_{\tilde{\Omega}} \tilde{L} d^n\tilde{x}$$ is the
transformed form of the functional $$\hat{{\mathcal L}}=\int_{\Omega} \hat{L}
d^nx$$ where $\hat{L}$ and $\tilde{L}$ are related as above, $\Omega$ is the
domain of integration and $\tilde{\Omega}$ is the transformed domain under
$j\bar{\psi}$ (see \cite{O86} pp.249-250). Notice that both of these are local
coordinate expressions of the equation $\displaystyle {\mathcal L} = \int_M L
\nu$, for appropriately restricted charts. Now suppose that $\psi$ is an
automorphism of $E$, $j\psi$ its induced automorphism on $J^\infty
E$, and $\psi _M$ its induced (orientation-preserving) diffeomorphism on 
$M$. Also suppose that
$\hat{L}$ and $\tilde{L}$ are two Lagrangians related by the equation
$(\tilde{L} \circ j\psi)\textup{det}(\psi _M) = \hat{L}$. We have:

\begin{lemma} \label{Lagrangian} Let $P$ be a Lagrangian as above, then
\begin{equation}
\label{ELI}
\E_a((P\circ j\psi)\textup{det}(\psi _M)) =\textup{det}(\psi _M) \frac{\partial
\psi^c_E}
{\partial u^a} (\tilde{\E}_c(P)\circ j\psi). \end{equation}
\end{lemma}

\begin{proof} First notice that
$\displaystyle \E_{u^a}(\hat{L}) = \textup{det}(\psi _M)
\frac{\partial \psi^c_E} {\partial u^a} (\E_{\tilde{u}^c}
(\tilde{L}) \circ j\psi)$ (see \cite{O86} pp.250).
But $(\tilde{L} \circ j\psi)\textup{det}(\psi _M) = \hat{L}$. The
identity $\ref {ELI}$
follows by letting $P=\tilde{L}$. {\em Notice that this is justified
since $\tilde{L}$ is arbitrary in the sense that given any $L'$ there
exists
an $\hat{L}$ derived from a Lagrangian
$L$ as above such that $(L' \circ j\psi)
\textup{det}(\psi _M) = \hat{L}$ since $j\psi$ is an automorphism}. \qed
\end{proof}

Let $\hat{\psi}$ denote the mapping representing the induced action of the
automorphism on sections of $E$, i.e., $\hat{\psi}: \Gamma E \rightarrow \Gamma
E$ where $\hat{\psi}(\phi) = \psi \circ \phi \circ \psi ^{-1}_M$ and $\phi$ is
a section of $E$. This induces a mapping on the space of local functionals
given by

\begin{eqnarray*}
\Psi(\mathcal P)=(\mathcal P \circ \hat{\psi})(\phi)
& = &
\mathcal P (\psi \circ \phi \circ \psi ^{-1}_M) \\ & = &
\int_M [P \circ j(\psi \circ \phi \circ \psi ^{-1}_M)] \nu \\ & = &
\int_M [P \circ j\psi \circ j\phi \circ \psi ^{-1}_M)] \nu \\ & = &
\int_M [P \circ j\psi \circ j\phi] (\textup{det} \psi _M) \nu, \\
\end{eqnarray*}
where $$\mathcal P (\phi)=\int_M (P \circ j\phi) \nu,$$ and $\phi$ is a
section
of $E$. \\

Now we find conditions on those automorphisms of the space of local functionals under which the Poisson structure is
preserved for a few special cases of the bundle $E$ and differential operators $\mathcal D$. \\

\subsection{Case I: dim($M$)=1, dim($E$)=2; $\mathcal D=\omega D_x$}
We begin with the case where $M$ is 1-dimensional, $E$ is 2-dimensional and consider
first order differential operators of the form $\mathcal D=\omega D_x$ with $\omega \in Loc_E$. {\em For simplicity we will skip the tilde's in our notation}.
Recall that
$\displaystyle \{\mathcal P,\mathcal Q\} = \int_M \mathcal D(\E(Q))\E(P)
\nu,$
and hence
\begin{equation}
\{\mathcal P,\mathcal Q\}(\phi) = \int_M [\mathcal D(\E(Q))\E(P)\circ j\phi]
\nu.
\end{equation}
In this case the Poisson bracket is preserved:
$\{\Psi(\mathcal P),\Psi(\mathcal Q) \}(\phi)=
\Psi(\{\mathcal P,\mathcal Q\})(\phi)$
if and only if

$\displaystyle \int_M ([\omega D_x\E((Q \circ j\psi)\textup{det}\psi 
_M)\E((P
\circ
j\psi) \textup{det}\psi _M)] \circ j\phi) \nu = $ $$\int_M
([(\omega D_x\E(Q)\E(P))\circ j\psi \circ j\phi]\textup{det}\psi 
_M)\nu,$$

\noindent
but since this latter equation holds for all sections $\phi$ of $E$ it 
is
equivalent to
$$\omega D_x\E((Q \circ j\psi) \textup{det}\psi _M)\E((P \circ j\psi)
\textup{det}\psi _M) =
[(\omega D_x\E(Q)\E(P))\circ j\psi] \textup{det}\psi _M$$
up to a {\em divergence}.
Using Lemma $\ref{Lagrangian}$ this is equivalent to \\
$\displaystyle \omega D_x(\textup{det}\psi _M \frac{\partial 
\psi_E} {\partial u} \E(Q) \circ j\psi)\textup{det}\psi _M
\frac{\partial \psi_E}{\partial u}
(\E(P) \circ j\psi) = $
$$(\omega \circ j\psi) (D_x(\E(Q)) \circ j\psi)
(\E(P) \circ j\psi) \textup{det}\psi _M$$
or \\
$\displaystyle \omega [D_x(\textup{det}\psi _M \frac{\partial 
\psi_E} {\partial u}) (\E(Q) \circ j\psi) + \textup{det}\psi _M \frac{\partial 
\psi_E} {\partial u}(D_x (\E(Q)) \circ j\psi) D_x\psi _M]$
$$\frac{\partial \psi_E}{\partial u}
(\E(P) \circ j\psi) =
(\omega \circ j\psi) (D_x(\E(Q)) \circ j\psi)
(\E(P) \circ j\psi)$$ \\
up to a divergence and where we have used equation ($\ref {der}$). (Notice that $\textup{det}\psi _M$ cancelled from both sides of the equation since it is nonzero.) Finally since the last equation is true for all $P$ and $Q$ we must have
$\displaystyle D_x(\textup{det}\psi _M \frac{\partial \psi_E} {\partial u})=0$ and
$\displaystyle \omega\circ j\psi = \omega (\textup{det}\psi _M)^2
(\frac{\partial \psi_E}{\partial u})^2.$ (One may factor out $(D_x (\E(Q)) \circ j\psi)$ over appropriate domains from the left-hand side of the equation to verify that we must have $\displaystyle D_x(\textup{det}\psi _M \frac{\partial \psi_E}{\partial u}) = 0$ since $Q$ is arbitrary, otherwise the equation may not hold for all $P$ and $Q$. Also notice that by the arbitrariness of $P$ and $Q$ the divergence term must be zero.) Observe that the first condition implies that 
$\displaystyle \textup{det}\psi _M \frac{\partial \psi_E} {\partial u}$ is a constant, while the second condition was simplified using $D_x \psi_M = \textup{det}\psi _M$ since $M$ is one-dimensional and so are its fibers. \\

\begin{theorem} Let $E\to M$ be a vector bundle with dim($M$)=1, dim($E$)=2 and let the Poisson bracket $\{\cdot,\cdot\}$ be defined by $\mathcal D=\omega D_x$. Then the induced transformation $\Psi$ on the space of local functionals is canonical, i.e.,
$\{\Psi(\mathcal P),\Psi(\mathcal Q)\}=\Psi(\{\mathcal P,\mathcal Q\})$ for all $\mathcal P, \mathcal Q \in \mathcal F$ if and only if \\
(i) $\displaystyle D_x(\textup{det}\psi _M \frac{\partial \psi_E} {\partial u})=0$, and \\
(ii) $\displaystyle \omega\circ j\psi = \omega (\textup{det}\psi _M)^2
(\frac{\partial \psi_E}{\partial u})^2.$
\end{theorem}

\subsection{Case II: dim($M$)=$n$, dim($E$)=$m+n$; $\mathcal D=\omega^{abi} D_i$}
Suppose that $M$ is $n$-dimensional with fibers of dimension $m$, and consider
first order differential operators of the form $\mathcal D=\omega^{abi} D_i$. Preserving the Poisson bracket:
$\{\mathcal P\circ \hat{\psi},\mathcal Q \circ \hat{\psi} \}(\phi)=
(\{\mathcal P,\mathcal Q\} \circ \hat{\psi})(\phi)$
is equivalent to

$\displaystyle \int_M ([\omega^{abi} D_i\E_b((Q \circ j\psi)\textup{det}\psi 
_M)\E_a((P
\circ
j\psi) \textup{det}\psi _M)] \circ j\phi) \nu = $ $$\int_M
([(\omega^{abi} D_i\E_b(Q)\E_a(P))\circ j\psi \circ j\phi]\textup{det}\psi 
_M)\nu,$$

\noindent
but since this latter equation holds for all sections $\phi$ of $E$ it 
is
equivalent to
$$\omega^{abi} D_i\E_b((Q \circ j\psi) \textup{det}\psi _M)\E_a((P \circ j\psi)
\textup{det}\psi _M) =
[(\omega^{abi} D_i\E_b(Q)\E_a(P))\circ j\psi] \textup{det}\psi _M$$
up to a {\em divergence}.
Using Lemma $\ref{Lagrangian}$ this is equivalent to \\
$\displaystyle \omega^{abi} D_i(\textup{det}\psi _M \frac{\partial 
\psi_E^d} {\partial u^b} \E_d(Q) \circ j\psi)\textup{det}\psi _M
\frac{\partial \psi_E^c}{\partial u^a}
(\E_c(P) \circ j\psi) = $
$$(\omega^{abi} \circ j\psi) (D_i(\E_b(Q)) \circ j\psi)
(\E_a(P) \circ j\psi) \textup{det}\psi _M$$
or \\
$\displaystyle \omega^{abi} [D_i(\textup{det}\psi _M \frac{\partial 
\psi_E^d} {\partial u^b}) (\E_d(Q) \circ j\psi) + \textup{det}\psi _M \frac{\partial 
\psi_E^d} {\partial u^b}(D_j \E_d(Q) \circ j\psi) D_i\psi _M^j]$
$$\frac{\partial \psi_E^c}{\partial u^a}
(\E_c(P) \circ j\psi) =
(\omega^{abi} \circ j\psi) (D_i(\E_b(Q)) \circ j\psi)
(\E_a(P) \circ j\psi)$$ \\
up to a divergence and where we have used equation ($\ref {der}$). Finally this can be rewritten as \\
$\displaystyle \omega^{cdj} [D_j(\textup{det}\psi _M \frac{\partial 
\psi_E^b} {\partial u^d}) (\E_b(Q) \circ j\psi) + \textup{det}\psi _M \frac{\partial 
\psi_E^b} {\partial u^d}(D_i (\E_b(Q)) \circ j\psi) D_j\psi _M^i]$
$$\frac{\partial \psi_E^a}{\partial u^c}
(\E_a(P) \circ j\psi) =
(\omega^{abi} \circ j\psi) (D_i(\E_b(Q)) \circ j\psi)
(\E_a(P) \circ j\psi)$$ up to a divergence.
As before for this to hold for all $P$ and $Q$ we must have $\displaystyle D_j(\textup{det}\psi _M \frac{\partial 
\psi_E^b} {\partial u^d})=0$ for all $d,j$ satisfying $\omega^{cdj} \neq 0$ for some $c$, and
$\displaystyle \omega^{abi} \circ j\psi=\textup{det}\psi _M\omega^{cdj} \frac{\partial \psi_E^a}{\partial u^c}\frac{\partial \psi_E^b} {\partial u^d}D_j\psi _M^i$. \\

\begin{theorem} Let $E\to M$ be a vector bundle with dim(M)=n, dim(E)= n+m, and let the Poisson bracket $\{\cdot,\cdot\}$ be defined by $\mathcal D=\omega^{abi} D_i$. Then the induced transformation $\Psi$ on the space of local functionals is canonical, i.e., $\{\Psi(\mathcal P),\Psi(\mathcal Q)\}=\Psi(\{\mathcal P,\mathcal Q\})$ for all $\mathcal P, \mathcal Q \in \mathcal F$ if and only if \\
(i) $\omega^{cdj} \neq 0$ for some c $\displaystyle \Rightarrow D_j(\textup{det}\psi _M \frac{\partial \psi_E^b} {\partial u^d})=0$ for all b, and \\
(ii) $\displaystyle \omega^{abi} \circ j\psi=\textup{det}\psi _M\omega^{cdj} \frac{\partial \psi_E^a}{\partial u^c}\frac{\partial \psi_E^b} {\partial u^d}D_j\psi _M^i$.
\end{theorem}

\subsection{Case III: dim($M$)=1, dim($E$)=2; $\mathcal D=\omega ^I D_I$}
Lastly we consider the case where we have a one-dimensional manifold $M$ with one-dimensional fibers and $n$-th order
differential operators of the form $\mathcal D=\omega ^I D_I= \omega ^n D_n+\omega ^{n-1} D_{n-1}+\cdots+\omega ^1 D_1+\omega ^0$. For simplicity we assume that $I$ is a {\em number} rather than a multi-index since in this case the base manifold is one-dimensional so, for example, $D_n$ means $(D_x)^n$ and $D_I=(D_x)^I$. 
Following similar analysis as before and skipping a few steps the Poisson bracket is preserved if and only if

$\displaystyle [\omega ^0 (\textup{det}\psi _M \frac{\partial 
\psi_E} {\partial u} \E(Q) \circ j\psi)+ \omega ^I D_I(\textup{det}\psi _M \frac{\partial 
\psi_E} {\partial u} \E(Q) \circ j\psi)] \textup{det}\psi _M$
$$ \frac{\partial \psi_E}{\partial u} (\E(P) \circ j\psi) = 
[(\omega ^0 \E(Q) + \omega ^I D_I \E(Q)) \E(P)] \circ j\psi \textup{det}\psi _M$$
or \\
$\displaystyle [\omega ^0 (\textup{det}\psi _M \frac{\partial 
\psi_E} {\partial u} \E(Q) \circ j\psi)+ \left ( \begin{array}{c} I \\ k \end{array}\right ) 
\omega ^I[D_k(\textup{det}\psi _M \frac{\partial 
\psi_E} {\partial u}) D_{I-k}(\E(Q) \circ j\psi)]$
$$\frac{\partial \psi_E}{\partial u}
(\E(P) \circ j\psi) =
[(\omega ^0 \E(Q) + \omega ^I D_I \E(Q))
\E(P)] \circ j\psi$$ \\
up to a divergence. Now observe that using equation $(\ref{der})$ we have
\begin{eqnarray*} D_I(F\circ j\psi) &=& (D_IF \circ j\psi) (\textup{det}\psi _M)^I
+ (D_{I-1}F\circ j\psi) (I,I-1) + \\ & & (D_{I-2}F\circ j\psi) (I,I-2) + \cdots + (D_1F\circ j\psi) D_{I-1}(\textup{det}\psi _M)\end{eqnarray*}
where $D_I=(D_x)^I$ and we use $(I,j)$ for the $\left ( \begin{array}{c} I-1 \\ j-1 
\end{array}\right )$ permutations of $j$ det$\psi _M$'s and $I-j$ $D_x$'s with det$\psi _M$ in the 
rightmost slot (this may be derived by induction). For example $(3,2)=D_x((\textup{det}\psi _M)^2) + \textup{det}\psi _M D_x (\textup{det}\psi _M)$. This yields the conditions we must have for a canonical transformation: \\
$(i) \displaystyle \omega ^0\circ j\psi = \textup{det}\psi _M
\omega ^0 (\frac{\partial \psi_E}{\partial u})^2.$ \\
$(ii) \displaystyle \omega ^I\circ j\psi = \frac{\partial \psi_E}{\partial u}\sum _{J=I}^n \sum _{j=I}^J \left ( \begin{array}{c} J \\ j \end{array}\right )
\omega ^J D_{J-j}(\textup{det}\psi _M \frac{\partial \psi_E}{\partial u})(j,I); I=1,2,\cdots,n.$ \\

Notice that we can combine condition $(i)$ with condition $(ii)$ if we let (0,0)=1 and $(j,0)=0$ for $j\neq 0$. To summarize

\begin{theorem} Let $E\to M$ be a vector bundle with dim(M)=1, dim(E)= 2, and let the Poisson bracket $\{\cdot,\cdot\}$ be defined by the n-th order differential operator $\mathcal D=\omega ^ID_I$. Then the induced transformation $\Psi$ on the space of local functionals is canonical, i.e., $\{\Psi(\mathcal P),\Psi(\mathcal Q)\}=\Psi(\{\mathcal P,\mathcal Q\})$ 
for all $\mathcal P, \mathcal Q \in \mathcal F$ if and only if $$
\displaystyle \omega ^I\circ j\psi = \frac{\partial \psi_E}{\partial u}\sum _{J=I}^n \sum _{j=I}^J \left ( \begin{array}{c} J \\ j \end{array}\right )
\omega ^J D_{J-j}(\textup{det}\psi _M \frac{\partial \psi_E}{\partial u})(j,I); I=0,1,\cdots,n.$$
\end{theorem}

\noindent
{\bf Remark} If $\textup{det}\psi _M$ and $\displaystyle \frac{\partial \psi_E}{\partial u}$ are constant then the
above condition reduces to: $\displaystyle \omega ^I\circ j\psi = \omega ^I (\textup{det}\psi _M)^{I+1}
(\frac{\partial \psi_E}{\partial u})^2; I=0,1,\cdots,n$.

\section{Transformation of evolutionary systems and their conservation laws}
Recall that an evolutionary system of equations takes the form $$u_t=\mathcal D \delta \mathcal H$$
where $\mathcal D$ is a Hamiltonian differential operator and $\mathcal H$ the 
Hamiltonian of the system. The variational derivative $\delta$ of a local functional
$\mathcal P = \int_M P \nu$ has the same components as $\E(P)$, the Euler-Lagrange operator applied to $P$ (we use $\E(P)$ rather than $\E(P\nu)$ for simplicity).
In the event of a canonical transformation we have $\{\Psi(\mathcal P),\Psi(\mathcal Q)\}=\Psi(\{\mathcal P,\mathcal Q\})$. 
Therefore if $\mathcal P$ is a conservation law for the evolutionary system $u_t = \mathcal D \delta{\mathcal H}$ with Hamiltonian ${\mathcal H}$, then $\Psi(\mathcal P)$ is a conservation law for the evolutionary system $u_t = \mathcal D \delta\Psi(\mathcal H)$ with Hamiltonian $\Psi({\mathcal H})$. \\

This transforms the evolutionary system and its conservation laws. We illustrate with a few examples.
The first example is somewhat trivial but we find it a good point to start as an illustration of our approach.
Also recall that the conditions for a canonical transformation are rather restrictive. In the second example, 
even though the action is orientation-reversing our conclusions will still hold. The same applies for Example 3. \\

\noindent
{\bf Example 1}\hspace{0.5mm} Consider the Korteweg-de Vries equation $u_t=u_{xxx}+uu_x$ with the transformation $\psi$ defined by $\displaystyle \tilde u=ku, \tilde x=\frac{x}{k}, k > 0$. Even though this transformation is somewhat trivial we find it useful to consider in this example. The Hamiltonian and Hamiltonian differential operator associated with the KdV equation can be chosen to be $\displaystyle {\mathcal H}=\int (-\frac{1}{2}u_x^2 + \frac{1}{6}u^3)dx$ and ${\mathcal D}=D_x$ respectively. Now
observe that $\displaystyle\textup{det}\psi _M \frac{\partial \psi_E} {\partial u}=1$ is a constant, and $\displaystyle \omega\circ j\psi = \omega (\textup{det}\psi _M)^2 (\frac{\partial \psi_E}{\partial u})^2 = 1$. So $\psi$ defines a canonical transformation of the Poisson bracket defined by ${\mathcal D}=D_x$. This transformation transforms our Hamiltonian to $\displaystyle {\mathcal H}'=\int (-\frac{1}{2k^4}u_x^2 + \frac{1}{6k^3}u^3)dx$ and consequently the KdV equation to $\displaystyle u_t=\frac{1}{k^3}u_{xxx}+\frac{1}{k^2}uu_x$. \\
Notice that the conservation laws $\displaystyle {\mathcal P_1}=\int _M \frac{1}{2}u^2 dx, {\mathcal P_2}=\int _M xu+ \frac{1}{2}tu^2 dx$, and $\displaystyle {\mathcal M}=\int _M u dx$ are trivially transformed to $\displaystyle \Psi({\mathcal P_1})=\int _M \frac{k}{2}u^2 dx, \Psi({\mathcal P_2})=\int _M (\frac{xu}{k}+ \frac{k}{2}tu^2) dx$, and $\displaystyle \Psi({\mathcal M})=\int _M u dx$. \\ \\

\noindent
{\bf Example 2}\hspace{0.5mm} Consider the Korteweg-de Vries equation $$u_t=u_{xxx}+uu_x$$ as above but this time with the transformation defined by $\displaystyle \tilde u=x^2u, \tilde x=\frac{1}{x}$. Let $M={\bf R}^+$ and $E=M\times {\bf R}^2$.
In this case $\tilde u_{\tilde x}=(x^2u_x + 2xu)(-x^2)$ and the new Hamiltonian is given by $\displaystyle {\mathcal H}'=\Psi({\mathcal H})=\int (\frac{1}{2}x^6u_x^2 + 2x^5uu_x+2x^4u^2-\frac{1}{6}x^4u^3)dx$. So the KdV equation is transformed to $u_t=D_x\E(H')$ where $H'= \frac{1}{2}x^6u_x^2 + 2x^5uu_x+2x^4u^2-\frac{1}{6}x^4u^3$, i.e., we have $$u_t= -24x^3u -2x^3u^2-x^4uu_x-36x^4u_x-12x^5u_{xx}-x^6u_{xxx}.$$ The conservation laws $\displaystyle {\mathcal P_1}=\int _M \frac{1}{2}u^2 dx, {\mathcal P_2}=\int _M xu+ \frac{1}{2}tu^2 dx$, and $\displaystyle {\mathcal M}=\int _M u dx$ are transformed to $\displaystyle \Psi({\mathcal P_1})=\int _M -\frac{1}{2}x^2u^2 dx, \Psi({\mathcal P_2})=\int _M -\frac{u}{x}+ \frac{k}{2}tu^2 dx$, and $\displaystyle \Psi({\mathcal M})=\int _M u dx$. \\ \\

\noindent
{\bf Example 3}\hspace{0.5mm} The Boussinesq equation can be written as a system \[ \begin{array}{ccc} u_t &= &v_x, \\
v_t &= &\frac{8}{3}uu_x+\frac{1}{3}u_{xxx}. \end{array} \] It has Hamiltonian $\displaystyle {\mathcal H}=\int _M(-\frac{1}{6} u_x^2+\frac{4}{9}u^3 + \frac{1}{2}v^2)dx$ with associated Hamiltonian differential operator \[ {\mathcal D} = \left ( \begin{array}{cc} 0 & D_x \\ D_x & 0 \end{array} \right ). \]
Let $M={\bf R}^+$ and $E=M\times {\bf R}^2$.
Now consider the transformation $\displaystyle \tilde u=x^2u, \tilde
v = x^2v, \tilde x=\frac{1}{x}$. The Hamiltonian ${\mathcal H}$ is transformed to 
$\displaystyle {\mathcal H}'=\Psi({\mathcal H})=\int _M(-\frac{1}{6}
(-2x^3u-x^4u_x)^2+\frac{4}{9}x^6u^3 +
\frac{1}{2}x^4v^2)\frac{dx}{x^2}$ =
$\displaystyle \int _M(-\frac{1}{6}
(4x^4u^2+4x^5uu_x+x^6u_x^2)+\frac{4}{9}x^4u^3 +
\frac{1}{2}x^2v^2)dx$. This transforms the system to
\[ \begin{array}{ccc} u_t &= &-2xv-x^2v_x, \\
v_t &= &
-8x^3u-\frac{163}{3}x^4u_x-\frac{62}{3}x^5u_{xx}-
\frac{16}{3}x^4u^2-
\frac{8}{3}x^4uu_x-2x^6u_{xxx}. \end{array} \]
The Boussinesq equation has the conservation law
$${\mathcal P}=\int_M
(\frac{1}{6}u_{xx}^2-2uu_x^2+\frac{8}{9}u^4+2uv^2-
\frac{1}{2}v_x^2)dx.$$ This is
transformed to $\displaystyle \Psi({\mathcal P})= \int_M
(6x^6u^2+6x^8u_x^2+\frac{1}{6}x^{10}u_{xx}^2+12x^7uu_x+2x^8uu_{xx}+
2x^9u_xu_{xx}-8x^6u^3
+8x^7u^2u_x+2x^8uu_x^2+\frac{8}{9}x^6u^4+8x^6uv^2+8x^7uvv_x+
2x^8uv_x^2-2x^4v^2-2x^5vv_x-
\frac{1}{2}x^6v_x^2)dx.$

\noindent
{\bf Acknowledgements} I would like to thank Professor Ron Fulp for useful remarks and corrections.

\end{document}